\documentclass[10pt,a4paper]{article}
\usepackage[T1]{fontenc}
\usepackage[utf8]{inputenc}
\usepackage{lmodern}
\usepackage[margin=21mm,top=22mm,bottom=22mm,columnsep=6mm]{geometry}
\usepackage{microtype}
\usepackage{amsmath,amssymb}
\usepackage{graphicx}
\usepackage{booktabs,longtable,array}
\usepackage[numbers,sort&compress]{natbib}

\usepackage[font=small,labelfont=bf,labelsep=period]{caption}
\usepackage{placeins}
\usepackage{xurl}
\usepackage[hidelinks]{hyperref}
\newcolumntype{L}[1]{>{\raggedright\arraybackslash\footnotesize}p{#1}}

\hypersetup{
  pdftitle={Trapping and Fracture of Particles in Three-Sphere Ball-Milling Collisions},
  pdfauthor={Ziyi Cui, Qiong Zhang, Suwei Liu, Richard M. Lueptow},
  pdfsubject={Granular contact mechanics and fracture in ball milling},
  pdfkeywords={ball milling, discrete element method, fracture, trapping condition, Weibull theory, Christensen criterion}
}
\begin{document}
\twocolumn[{
\begin{center}
{\LARGE\bfseries Trapping and Fracture of Particles in Three-Sphere Ball-Milling Collisions\par}
\vspace{12pt}
{\large
\href{https://orcid.org/0009-0006-8899-6974}{Ziyi Cui}\textsuperscript{1},
\href{https://orcid.org/0000-0002-3856-6902}{Qiong Zhang}\textsuperscript{2},
\href{https://orcid.org/0000-0002-3592-6570}{Suwei Liu}\textsuperscript{2}, and
\href{https://orcid.org/0000-0002-1855-1471}{Richard M. Lueptow}\textsuperscript{2,3,4,*}\par}
\vspace{8pt}
{\small
\textsuperscript{1}University of Toronto Schools, Toronto, Ontario M5S 2R7, Canada\par
\textsuperscript{2}Department of Mechanical Engineering, Northwestern University, Evanston, Illinois 60208, USA\par
\textsuperscript{3}Department of Chemical and Biological Engineering, Northwestern University, Evanston, Illinois 60208, USA\par
\textsuperscript{4}Northwestern Institute on Complex Systems (NICO), Northwestern University, Evanston, Illinois 60208, USA\par
\vspace{4pt}
\textsuperscript{*}Corresponding author: \href{mailto:r-lueptow@northwestern.edu}{r-lueptow@northwestern.edu}\par}
\end{center}
\vspace{4pt}
\begin{minipage}{\textwidth}
\small
\textbf{Abstract.} We numerically investigate the collision and fracture processes of ball milling based on a minimal model of a three-sphere geometry using discrete element method (DEM) simulations combined with Christensen-type and Weibull-type normalized fracture criteria.
The outcome of the brittle target particle (ejected versus trapped, fractured versus intact) depends on the impact geometry and velocity.
We develop a reduced three-sphere collision description in which trapping is governed by impact geometry and friction, while fracture inside the trapping regime is evaluated from DEM output using stress-based and contact-force-based normalized criteria.
The trapping regime follows a closed-form geometric condition, while the critical conditions for fracture vary with impact angle, obliquity angle, impact velocity, and particle size ratio. Head-on collisions with larger size ratios require lower critical impact velocities for fracture.

\end{minipage}
\vspace{16pt}
}]
\section{Introduction}

Grinding is a critical process applied in a wide range of industries, including mineral processing~\cite{abouzeid2009grinding}, cement~\cite{aydougan2006high}, ceramics~\cite{malkin1996grinding}, food~\cite{gao2020recent}, pharmaceuticals~\cite{nakach2004comparison}, and the beneficiation of rare-earth metals~\cite{ji2021effect}.
Despite its broad application, ball milling is among the least efficient industrial operations~\cite{tavares2017review}. Comminution alone consumes up to 4\% of global industrial electricity~\cite{jeswiet2016energy} and approximately 36\% of a mine's total power usage~\cite{ballantyne2014benchmarking}, yet less than 1\% of the input energy actually produces new surface area, with the remainder dissipated as heat, noise, and vibration~\cite{fuerstenau2002energy,alvarado1998energy,zhang2022energy}. This inefficiency raises operating costs and magnifies environmental impacts, including increased carbon emissions~\cite{ige2024carbon,norgate2010energy,kittipongvises2017assessment}. Because grinding is so widely adopted, even modest efficiency improvements would translate into substantial energy and cost savings while reducing greenhouse gas emissions.

Among his many contributions to the field of particle and powder technology, Mojtaba Ghadiri has studied various aspects of milling operations and particle mechanics~\cite{GHADIRI20005445,ghadiri2002impact,ghadiri20051441,ghadiri20114757,ghadiri2011aspirin,ghadiri2015milltype,ghadiri2021239,ghadiri2021204,ghadiri2025120822}.
To connect the physics of individual collisions with large-scale milling, previous studies have used experiments to measure breakage. Single-particle indentation~\cite{ghadiri2011aspirin}, impact~\cite{tavares1998single,ghadiri2002impact}, and drop-weight experiments~\cite{genc2004single}, and pilot-scale mill trials~\cite{erdem2009effect} measure the strength and probability of breakage under controlled loading. These studies reveal real material behavior but cannot practically span the wide range of impact geometries, size ratios, and frictional conditions seen in industrial mills, and scaled-up experiments are costly, time-intensive, and often difficult to reproduce. Theoretical approaches such as population balance models describe particle size distribution changes across milling times and are computationally efficient~\cite{zhang2022energy}, but they rely heavily on empirical fitting and do not resolve the mechanics of individual collisions, which are the actual triggers of fracture~\cite{ramkrishna1985status}. In short, experiments provide realism but narrow scope, while population balance models provide generality but limited physical detail.

Numerical simulation offers a way to overcome some of these limitations. The discrete element method (DEM) has become one of the most versatile tools for studying granular materials~\cite{cundall1979discrete,ketterhagen2009process,toledo2025review} because it explicitly resolves particle trajectories, contact forces, and energy dissipation by treating each particle as a discrete entity and integrating its motion under Newton's laws. Unlike population balance models, DEM does not require empirical fitting to capture local collision mechanics; unlike experiments, DEM can systematically sweep impact velocity, angles, size ratio, and friction over a range of values that is difficult to achieve in the laboratory. At the same time, simulation outcomes are most powerful when interpreted together with theoretical models that encode the governing physics, so that DEM and theory together provide both predictive capability and particle-level insight.

This study develops a simulation framework in which the collision outcome is separated into two stages that can be analyzed independently. The first stage is geometric and frictional: whether the target particle is trapped between a pair of grinding balls \cite{PRZIWARA2020evaluation} or is ejected before it can be loaded. The second stage is mechanics-based: whether a trapped particle experiences sufficient compressive loading to fracture. A three-sphere collision system is used as a minimal model to represent the local environment of a ball mill while allowing precise control of impact velocity $v$, impact angle $\alpha$, obliquity angle $\gamma$, size ratio $R$, and friction coefficient $\mu$. We show that trapping is governed by a closed-form geometric condition that generalizes the classical angle-of-nip criterion of roll-crusher and high-pressure grinding-roll theory to a three-sphere geometry \cite{Johanson1965rolling,bindhumadhavan2005roll,yehia2007estimation,kleinebudde2022improving}. We then evaluate fracture within the trapping regime with two failure criteria applied to the DEM output, one a global stress-based limit and the other a local, flaw-based criterion: a Christensen failure function based on stresses in the target particle and a Weibull failure function based on the contact forces experienced by the target particle. By combining DEM with analytic contact geometry, this study provides both detailed numerical results and a compact mechanistic picture that sheds light on improving grinding efficiency in a broad range of ball milling applications.

\section{Geometry of Impacts}
\label{sec:methods}\label{sec:geometry}

\begin{figure}[!htb]
  \centering
  \includegraphics[width=\linewidth]{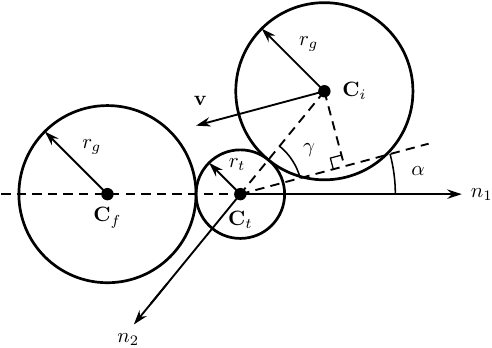}
  \caption{Schematic of the planar three-sphere collision at the instant of contact in the plane of the three spheres. A fixed grinding ball (center $\mathbf{C}_f$, radius $r_g$) is tangent to a smaller brittle target particle (center $\mathbf{C}_t$, radius $r_t$), while an incoming grinding ball (center $\mathbf{C}_i$, radius $r_g$) approaches with velocity $\mathbf{v}$.  The horizontal line through $\mathbf{C}_f$ and $\mathbf{C}_t$ defines the $n_1$ axis, and the dashed ray through $\mathbf{C}_t$ drawn parallel to $\mathbf{v}$ makes the impact angle $\alpha$ with $n_1$. The line through the centers $\mathbf{C}_t$ and $\mathbf{C}_i$ defines the $n_2$ axis and forms the obliquity angle $\gamma$ with the dashed ray parallel to $\mathbf{v}$. }
  \label{fig:geometric-setup}
\end{figure}

Fig.~\ref{fig:geometric-setup} shows the planar three-sphere collision studied here.  Two identical grinding balls have radius $r_g$, and the smaller target particle has radius $r_t$, resulting in a size ratio $R=r_g/r_t$. One grinding ball is fixed, so it can neither translate nor rotate, and centered at $\mathbf{C}_f$, initially tangent to the target particle at $\mathbf{C}_t$. The incoming grinding ball, centered at $\mathbf{C}_i$, begins a short distance away and travels with initial in-plane velocity $\mathbf{v}$ until it collides with the target particle.

Simultaneous contacts, $\mathbf{C}_f-\mathbf{C}_t$ and $\mathbf{C}_i-\mathbf{C}_t$, result in the contact forces $\mathbf{F}_1$ and $\mathbf{F}_2$ along the corresponding axes, the lines of centers: the $n_1$ axis passes through $\mathbf{C}_f$ and $\mathbf{C}_t$, and the $n_2$ axis through $\mathbf{C}_i$ and $\mathbf{C}_t$. The impact angle $\alpha$ is measured between $-\mathbf{v}$ and the $n_1$ axis ($\alpha \in [-90^\circ, 90^\circ]$), and the obliquity angle $\gamma$ is between $\mathbf{v}$ and the $n_2$ axis ($\gamma \in [0^\circ, 90^\circ)$). See Section~\ref{sec:si_nomenclature} in the Supplementary Information for the complete nomenclature.

\begin{figure}[!htbp]
  \centering

  \includegraphics[width=\linewidth]{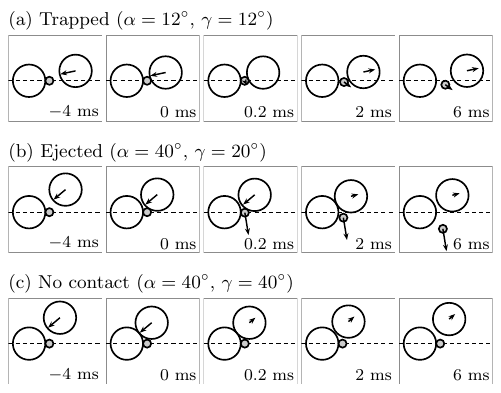}

  \caption{Snapshots based on DEM simulations for $v=3$~m/s, $R=4$ ($r_g=25$~mm, $r_t=6.25$~mm), and $\mu=0.47$ with three outcomes, depending on $(\alpha,\gamma)$: trapped ($\alpha=12^\circ$, $\gamma=12^\circ$), ejected ($\alpha=40^\circ$, $\gamma=20^\circ$), and no contact ($\alpha=40^\circ$, $\gamma=40^\circ$). Time is measured from the initial contact of the incoming grinding ball, which starts $30$~mm away, and continues for $7$~ms after first contact.  The fixed grinding ball (left) and incoming grinding ball are open circles, the target particle is the small shaded circle, and the dashed horizontal line is along the $n_1$ axis. The arrows are the velocity vectors, with length proportional to speed. }
  \label{fig:outcomes}
\end{figure}

Depending on the impact geometry, the target particle is either not contacted at all, ejected from between the two grinding balls, or trapped between them, as shown in Fig.~\ref{fig:outcomes} for $v = 3$~m/s and $R = 4$, with $\mu = 0.47$. In the trapped case, the target particle is captured between the incoming grinding ball and fixed grinding ball so that it is compressed until the incoming grinding ball rebounds (Fig.~\ref{fig:outcomes}(a)). If the compressive stress is high enough, the target particle will be crushed and break up into smaller pieces. At lower compressive stresses, the target particle will remain intact and is released, as shown in the figure.  Changing the impact and obliquity angles so that the incoming grinding ball hits the target particle less directly results in the target particle being ejected as it is squeezed out between the two balls (Fig.~\ref{fig:outcomes}(b)). Although the target may be compressed slightly, it is unlikely that it will be crushed. The ejection depends on the friction at the contact points between the grinding balls and the target particles. At low friction, the target particle easily slips out and is ejected, and the grinding balls themselves collide, but at high enough friction the target particle is trapped rather than ejected. Finally, for certain conditions, the incoming grinding ball misses the target particle altogether and impacts only the fixed grinding ball (Fig.~\ref{fig:outcomes}(c)). These three conditions can be predicted on the basis of the geometry and contact forces alone.

The initial configuration $(v,\alpha,\gamma,R,\mu)$ determines the outcome. The incoming grinding ball must first reach the target particle, and then the target particle must be trapped between the two grinding balls rather than ejected. Later in this paper, we consider situations where the impact velocity exceeds a critical value necessary to fracture the target particle. Consider the first two conditions in sequence.

\subsection{Contact Condition}
\label{sec:impact}

\begin{figure}[!htbp]
  \centering

  \makebox[0.75\linewidth][l]{(a)}\par\vspace{-0.3em}
  \includegraphics[width=0.75\linewidth]{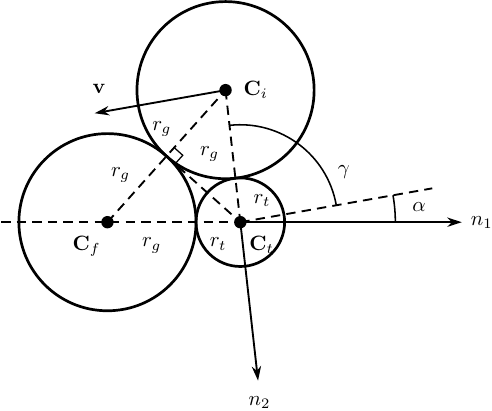}\par\medskip
  \makebox[0.75\linewidth][l]{(b)}\par\vspace{-0.3em}
  \includegraphics[width=0.6928\linewidth]{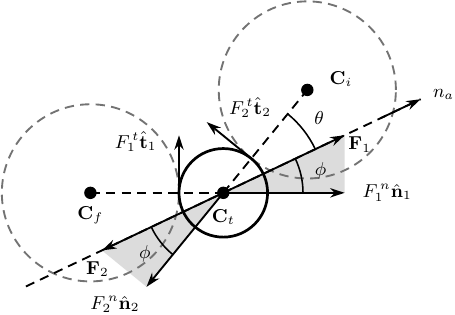}
  \caption{
(a) Limiting contact geometry at which the incoming grinding ball (center $\mathbf{C}_i$, velocity $\mathbf{v}$) touches the fixed grinding ball ($\mathbf{C}_f$) and the target particle ($\mathbf{C}_t$) at the same instant, shown for $R=2$.  (b) Free-body diagram of the target particle at the critical trapping case $\alpha+\gamma=2\phi$, where $\phi$ is the half-angle of the Coulomb friction cone defined by Equation~\eqref{eq:phi}. The contact forces are normal components $F_i^{\,n}\hat{\mathbf{n}}_i$ and tangential components $F_i^{\,t}\hat{\mathbf{t}}_i$, with resultants $\mathbf{F}_i$ (in the case shown $F_2^{\,t}<0$, so the arrow labeled $F_2^{\,t}\hat{\mathbf{t}}_2$ points along $-\hat{\mathbf{t}}_2$); the shaded triangles show the half of the Coulomb friction cone of Equation~\eqref{eq:phi} on the side of the friction force at each contact, within which the resultant contact force $\mathbf{F}_i$ must lie; $\theta$ is the half-angle, equal to $\phi$ in the case shown; and $n_a$ is the common line of action of the two resultants, used in Section~\ref{sec:christensenforce}. The grinding balls are drawn dashed.
}
  \label{fig:trapping-condition}
\end{figure}

Whether the incoming grinding ball can reach the target particle is determined solely by the opening angle $\alpha+\gamma$ and the size ratio, with the limiting case shown in Fig.~\ref{fig:trapping-condition}(a). Contact with the target occurs before contact between the two grinding balls only if
\begin{equation}
    180^\circ-|\alpha+\gamma| \geq 2\arcsin \frac{r_g}{r_g+r_t},
\end{equation}
or, equivalently, in terms of $R=r_g/r_t$
\begin{equation}
  |\alpha + \gamma| \le 2\arccos\frac{R}{R+1}\text{.}
  \label{eq:reach_limit}
\end{equation}

For larger values of $|\alpha+\gamma|$, the incoming grinding ball hits the fixed grinding ball before it hits the target particle, and the outcome is no contact between the incoming grinding ball and the target particle.

\subsection{Trapping Condition}
\label{sec:trapping}

Whether the target particle is trapped between the two grinding balls or ejected before it can be sufficiently loaded is determined by the collision geometry and the friction coefficient. As shown in Fig.~\ref{fig:trapping-condition}(b), the normal forces tend to squeeze the target particle out, while the frictional tangential forces resist this.

The construction described below generalizes the classical angle-of-nip condition of roll-crusher and high-pressure grinding-roll theory \cite{Johanson1965rolling,bindhumadhavan2005roll,yehia2007estimation,kleinebudde2022improving}, in which a particle is captured only if the half-angle subtended at the two loading surfaces does not exceed the friction angle, $\phi$. The classical criterion assumes a symmetric geometry with a single nip half-angle. Here the two contact normals are not symmetric about the approach direction, so the condition involves two independent angles and emerges in the combination $|\alpha+\gamma| \le 2\phi$; the symmetric roll geometry is recovered as the special case $\alpha=\gamma$.

The construction is summarized in Fig.~\ref{fig:trapping-condition}(b). The target particle experiences a normal force from each grinding ball, $F_1^{\,n}$ along $\hat{\mathbf{n}}_1$ from the fixed grinding ball and $F_2^{\,n}$ along $\hat{\mathbf{n}}_2$ from the incoming grinding ball, together with a tangential friction force $F_i^{\,t}$ at each contact along $\hat{\mathbf{t}}_1 = \hat{\mathbf{z}} \times \hat{\mathbf{n}}_1$ and $\hat{\mathbf{t}}_2 = -\hat{\mathbf{z}} \times \hat{\mathbf{n}}_2$. The resultant contact forces on the target particle are
\begin{equation}
  \mathbf{F}_1 = F_1^{\,n}\,\hat{\mathbf{n}}_1 + F_1^{\,t}\,\hat{\mathbf{t}}_1, \qquad
  \mathbf{F}_2 = F_2^{\,n}\,\hat{\mathbf{n}}_2 + F_2^{\,t}\,\hat{\mathbf{t}}_2.
\end{equation}
At each contact, the normal direction $\hat{\mathbf{n}}_i$ is fixed by geometry, whereas the static friction force can take any value satisfying $|F_i^{\,t}|\le \mu F_i^{\,n}$. Since $\mathbf{F}_i$ deviates from its contact normal by the angle $\arctan(|F_i^{\,t}|/F_i^{\,n})$, which cannot exceed $\arctan\mu$, the resultant contact force is not known a priori and may lie anywhere within a cone of half-angle
\begin{equation}
  \phi = \arctan\mu
  \label{eq:phi}
\end{equation}
about its contact normal.

Since the inertia of the target particle is negligible compared with that of the grinding balls, the quasi-static condition can be applied to the target particle, leading to the force balance $\mathbf{F}_1 = -\mathbf{F}_2$.
For the target particle to be held between the grinding balls rather than ejected sideways, the direction of $\mathbf{F}_1$ must lie within $\phi$ of $\hat{\mathbf{n}}_1$ and the direction of $\mathbf{F}_2$ must be within $\phi$ of $\hat{\mathbf{n}}_2$, as indicated by the shaded triangles in Fig.~\ref{fig:trapping-condition}(b). Since $\hat{\mathbf{n}}_1$ and $-\hat{\mathbf{n}}_2$ are separated by the angle $\alpha+\gamma$ (see Fig.~\ref{fig:geometric-setup}), define $\theta=\tfrac{1}{2}(\alpha+\gamma)$ as the corresponding half-angle, shown in Fig.~\ref{fig:trapping-condition}(b) between the resultant force direction (along $n_a$) and the line of centers $\mathbf{C}_t\mathbf{C}_i$. This trapping condition can be met only if a single resultant direction can lie within $\phi$ of both contact normals, i.e.\ only if
\begin{equation}
    |\theta|\leq\phi = \arctan\mu.\label{eq:trap_boundary}
\end{equation}
The equality, $\theta=\phi$, is the trapping boundary and corresponds to the critical case shown in Fig.~\ref{fig:trapping-condition}(b). The trapping condition therefore compares a purely geometric angle, $\theta=(\alpha+\gamma)/2$, set by the collision geometry through $\alpha$ and $\gamma$, with the half-angle of the Coulomb friction cone, $\phi=\arctan\mu$, set by the contacting materials through $\mu$.

The target particle is compressed under sustained double contact (trapped) only if both the contact condition of Equation~\eqref{eq:reach_limit} and the frictional condition of Equation~\eqref{eq:trap_boundary} are satisfied. Combining the frictional bound $|\alpha + \gamma| \le 2\arctan\mu$ with the contact condition of Equation~\eqref{eq:reach_limit} yields the trapping regime
\begin{equation}
  |\alpha + \gamma| \le \min\!\left(\,2\arctan\mu,\; 2\arccos\frac{R}{R+1}\,\right).
  \label{eq:trap_band}
\end{equation}
For larger values of $|\alpha+\gamma|$, the target particle is either ejected or there is no contact between the incoming grinding ball and the target particle at all.

\section{Simulation Methods}
\label{sec:dem}

With this background, we now simulate the simple collision of an incoming grinding ball on a stationary target particle already in contact with a fixed grinding ball (Fig.~\ref{fig:geometric-setup}) over the parameter ranges $v \in [0.5,5]~\mathrm{m/s}$ (varied in steps of 0.1~m/s), typical of ball mills~\cite{cleary2001charge,datta2002direct} (see Section~\ref{sec:si_ranges} in the Supplementary Information for details), $\alpha \in [-90^\circ,90^\circ]$ (steps of $4^\circ$), $\gamma \in [0^\circ,90^\circ)$ (steps of $2^\circ$), $R=r_g/r_t \in [2,10]$ (steps of $0.4$), and $\mu \in [0,1]$ (steps of $0.025$). The size ratio $R$ is varied by changing $r_t$ while holding $r_g = 25$~mm fixed. The three-sphere dynamics are simulated with the open-source LAMMPS package~\cite{thompson2022lammps}. Contacts are modeled using Hertz's model with a contact restitution coefficient \cite{antypov2011analytical} in the normal direction and the history-dependent Mindlin model \cite{mindlin1949compliance} limited by Coulomb sliding friction in the tangential direction. The tangential damping is disabled, so tangential dissipation occurs through Coulomb sliding alone; the tangential stiffness is derived from $E$ and $\nu$ rather than prescribed independently.

Corresponding to a common ball milling setup, we use as an example limestone for the target particle material and steel for the grinding balls, although the approach applies to any target particle material or grinding ball material.
Table~\ref{tab:materials} lists the material and contact properties \cite{grange2008dynamic,chen2016variation,wang2012investigation}, with subscripts $t$ and $g$ denoting the target particle and grinding balls, respectively.
Based on the same-material sliding friction coefficients in Table~\ref{tab:materials}, the limestone--steel contact uses the geometric average, $\mu=\sqrt{0.75\times 0.30}\approx 0.47$.
$\mu$ is varied later to examine the dependence of impact outcomes on friction.
No measured value of the limestone--steel restitution coefficient could be located in the literature, so $e = 0.50$ is adopted as a plausible value.

The equations of motion are integrated with the synchronized velocity-Verlet scheme \cite{vyas2025improved} with a timestep of $\Delta t=1.0\times10^{-7}~\mathrm{s}$, resolving each binary contact with hundreds of timesteps over the investigated material and size ranges. Output is recorded every $100$ timesteps; at this interval, the recovered force and stress peaks are converged to within $0.13\%$ (Section~\ref{sec:si_convergence} in the Supplementary Information). Each production run is integrated from launch, with the incoming grinding ball initialized with a 10~mm surface-to-surface gap and advanced at constant velocity until first contact. Time is advanced through the duration of the contact until the incoming grinding ball is no longer in contact with either of the other two spheres. Configurations in which the incoming grinding ball never reaches the target particle are recorded as no-contact events.

For cases where the incoming grinding ball collides with the target particle, we obtain the peak volume-averaged compressive principal stress, $\sigma_{\mathrm{peak}}$,  as follows. The volume-averaged stress tensor of the target particle, contributed from the contact forces, is defined as \cite{weinhart2013discrete,hill2014segregation}:
\begin{equation}
  \bar{\boldsymbol{\sigma}} \;=\; \frac{1}{V_t}\sum_{i}\mathbf{F}_i\otimes\boldsymbol{\ell}_i,
  \label{eq:avgstress}
\end{equation}
where $V_t = \tfrac{4}{3}\pi r_t^3$ is the target particle volume, $\otimes$ denotes the outer product, and $\mathbf{F}_i$ and $\boldsymbol{\ell}_i$ are the contact force and the branch vector from the target particle center to contact $i$, $\boldsymbol{\ell}_i=-r_t\hat{\mathbf{n}}_i$. Equation~\eqref{eq:avgstress} can be applied to both head-on and oblique collisions. However, LAMMPS reports the ``per-atom'' stress of the target particle from the per-atom virial tensor \cite{thompson2009general,lammps:compute_stress_atom}, which assigns to each contact $i$ one half of the pair contribution,
\begin{equation}
\bar{\boldsymbol{\sigma}}_{\mathrm{LAMMPS}}=\frac{1}{V_t}\sum_i \frac{1}{2}\,\mathbf{F}_i\otimes\mathbf{r}_i \text{,}
\label{eq:lammps_stress}
\end{equation}
where $\mathbf{r}_i$ is the center-to-center vector from the target particle to the grinding ball, of length $r_g+r_t$, and is collinear with the branch vector $\boldsymbol{\ell}_i=-r_t\hat{\mathbf{n}}_i$. Hence, $\mathbf{r}_i=\frac{r_g+r_t}{r_t}\boldsymbol{\ell}_i$ and the two definitions differ only by the corresponding length ratio. The volume-averaged stress tensor in Equation~\eqref{eq:avgstress} is therefore obtained from the LAMMPS output by rescaling every component as
\begin{equation}
\bar{\boldsymbol{\sigma}}=\frac{2r_t}{r_g+r_t}\,\bar{\boldsymbol{\sigma}}_{\mathrm{LAMMPS}}=\frac{2}{1+R}\,\bar{\boldsymbol{\sigma}}_{\mathrm{LAMMPS}} .
\label{eq:branch_rescale}
\end{equation}
We define $\sigma_{\mathrm{peak}}$ as the absolute value of the most compressive principal stress of $\bar{\boldsymbol{\sigma}}$, taken as the most negative eigenvalue, over the contact window. Note that this approach provides the peak volume-averaged compressive principal stress, but it does not resolve the local stress field inside the sphere, for which no closed-form solution exists even for diametral loading~\cite{pejchal2017compression}.

\begin{table*}[htbp]
  \centering  \rmfamily
  \caption{Material and contact properties used in the DEM simulations.}
  \label{tab:materials}
  \begin{tabular}{@{}lccc@{}}
    \toprule
    Property & Symbol & Target particle (limestone) & Grinding ball (steel) \\
    \midrule
    Density (kg\,m$^{-3}$)        & $\rho$   & 2700~\cite{grange2008dynamic}  & 7800~\cite{chen2016variation} \\
    Young's modulus (GPa)         & $E$      & 77.82~\cite{grange2008dynamic} & 208~\cite{chen2016variation}  \\
    Poisson's ratio               & $\nu$    & 0.28~\cite{grange2008dynamic}  & 0.27~\cite{wang2012investigation} \\

    Sliding friction coefficient  & $\mu_s$  & 0.75~\cite{wang2012investigation}  & 0.30~\cite{wang2012investigation} \\

    \bottomrule
  \end{tabular}
\end{table*}

\section{Simulation Results}
\label{sec:results}

The DEM simulations provide results for $\sigma_{\mathrm{peak}}$ over a broad range of impact velocities, collision geometries, friction, and size ratios.

\subsection{Effect of Collision Geometry on Stresses}
\label{sec:stress_geometry}

\begin{figure}[!htbp]
  \centering
  \includegraphics[width=\linewidth]{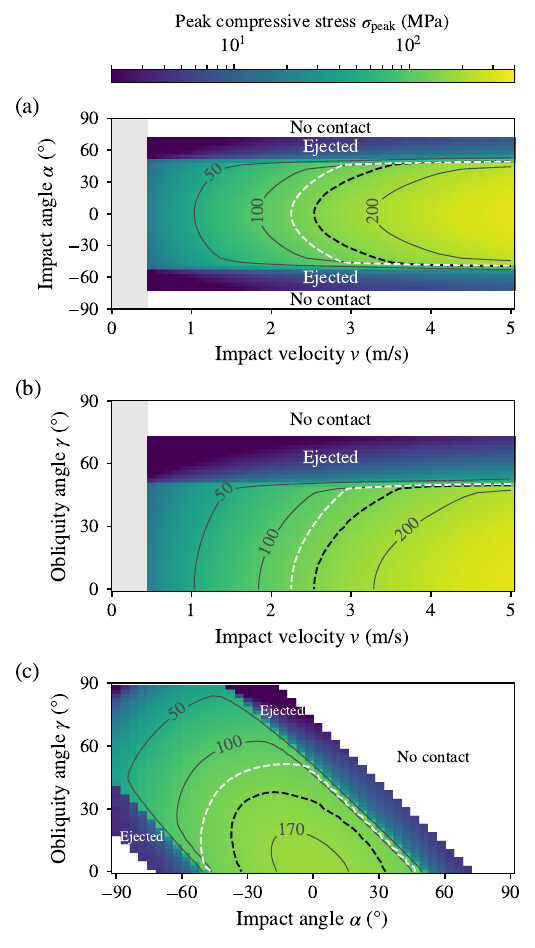}
  \caption{Stress maps at $R = 4$ and $\mu = 0.47$ with fracture boundaries overlaid. (a) $v$--$\alpha$ map at $\gamma = 0^\circ$. (b) $v$--$\gamma$ map at $\alpha = 0^\circ$. (c) $\alpha$--$\gamma$ map at $v = 3$~m/s. Color indicates the peak compressive principal stress in the target particle during the collision, $\sigma_{\mathrm{peak}}$, on a logarithmic scale with contours of constant $\sigma_{\mathrm{peak}}$ in MPa. Gray indicates that no simulations were performed. The dashed black curve indicates the fracture boundary for the Christensen criterion ($f_C=1$, Equation~\eqref{eq:fC}), and the dashed white curve indicates the fracture boundary for the Weibull criterion ($f_W=1$, Equation~\eqref{eq:fW}).}
  \label{fig:velocity-geometry-regimes}
\end{figure}

The effect of collision geometry on the peak compressive stress is shown for $R=4$ in Fig.~\ref{fig:velocity-geometry-regimes}, where the color encodes the peak compressive principal stress in the target particle during the collision, $\sigma_{\mathrm{peak}}$, on a logarithmic scale. Consider first Fig.~\ref{fig:velocity-geometry-regimes}(a), showing the dependence of $\sigma_{\mathrm{peak}}$ on impact velocity $v$ and impact angle $\alpha$ for situations where the velocity is along the line of impact, corresponding to obliquity angle $\gamma = 0^\circ$. It is evident that $\sigma_{\mathrm{peak}}$ increases with impact velocity, as would be expected, and is a maximum at $\alpha=0^\circ$, where the target particle is trapped directly between the grinding balls upon head-on impact. $\sigma_{\mathrm{peak}}$ drops off when $\alpha$ deviates from head-on impact as it becomes more positive or negative. $\sigma_{\mathrm{peak}}$ drops off sharply, regardless of impact velocity, for $|\alpha|$ exceeding $50.3^\circ$, which corresponds to the trapping boundary based on Equation~\eqref{eq:trap_boundary} for the target particle to be ejected, although the target particle still experiences some compressive stress. For $|\alpha|$ exceeding $73.7^\circ$, corresponding to the no-contact condition based on Equation~\eqref{eq:reach_limit}, $\sigma_{\mathrm{peak}}$ is negligible. Next, consider the effect of the obliquity angle $\gamma$ at $\alpha = 0^\circ$, shown in Fig.~\ref{fig:velocity-geometry-regimes}(b). The results are similar to those for varying $\alpha$ in Fig.~\ref{fig:velocity-geometry-regimes}(a): $\sigma_{\mathrm{peak}}$ is highest for head-on impact ($\gamma=0^\circ$) and decreases as the obliquity angle grows. In this case, the target particle is ejected for $\gamma \gtrsim 50.3^\circ$ and the incoming grinding ball misses the target particle altogether for $\gamma \gtrsim 73.7^\circ$.

To complete the picture, consider the dependence of $\sigma_{\mathrm{peak}}$ on both $\alpha$ and $\gamma$ at constant $v=3$~m/s in Fig.~\ref{fig:velocity-geometry-regimes}(c). Because of the impact geometry, the stress map is shifted toward negative $\alpha$ and positive $\gamma$. Here the trapping boundary of Equation~\eqref{eq:trap_boundary} corresponds to the two diagonal edges of the regions labeled ``Ejected'' at $|\alpha+\gamma|\approx 50.3^\circ$, and no contact occurs in the white region beyond $|\alpha + \gamma| \approx 73.7^\circ$.

The results in Fig.~\ref{fig:velocity-geometry-regimes} suggest that it may be possible to enhance the effectiveness of a milling operation by steering collisions into the trapping regime by operating at rotation speeds that favor cataracting motions and by lifter and liner designs that promote direct, near-head-on impacts~\cite{cleary2001charge,datta2002direct}, so that the loading inside the trapping regime, rather than ejection, decides the outcome.

\subsection{Effect of Friction on Stresses}
\label{sec:stress_friction}

\begin{figure}[!htb]
  \centering
  \includegraphics[width=\linewidth]{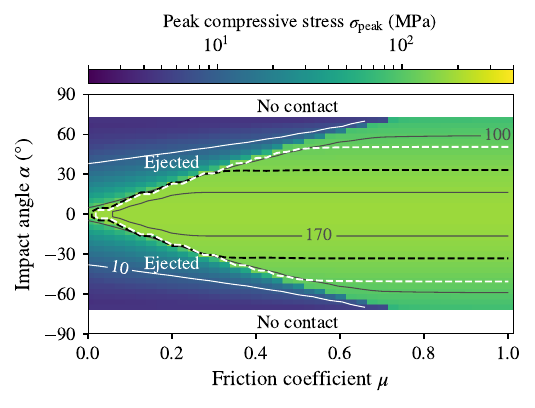}
  \caption{$\mu$--$\alpha$ map at $v = 3$~m/s, $\gamma = 0^\circ$, $R = 4$, with the same color and line conventions as Fig.~\ref{fig:velocity-geometry-regimes}.}
  \label{fig:friction-effect}
\end{figure}

A critical aspect evident in Fig.~\ref{fig:velocity-geometry-regimes} is the limitation resulting from target particle ejection, which depends on friction between the particles. The effect of friction on the peak compressive stress is examined in Fig.~\ref{fig:friction-effect} for $R=4$, $v = 3$~m/s, and $\gamma = 0$. The high-stress region broadens with increasing $\mu$ following the trapping boundary $\pm 2\arctan\mu$, according to Equation~\eqref{eq:trap_boundary}. At $\mu=0$, the target particle is ejected for any $\alpha\neq 0^\circ$, while for $\mu \ge 0.75$ the target particle is never ejected, although it is not impacted at all for $|\alpha| \ge 73.7^\circ$. Within the trapped region, $\sigma_{\mathrm{peak}}$ contours are independent of $\mu$ (horizontal), because the stress of a trapped particle depends on $\alpha$, $\gamma$, and $v$ but not on $\mu$, while $\gamma=0$ and $v=3$~m/s remain fixed in Fig.~\ref{fig:friction-effect}. For head-on collisions ($\alpha = 0$), $\sigma_{\mathrm{peak}}$ is independent of $\mu$ over the entire range $\mu \in [0,1]$, as seen by the uniform color along a horizontal line $\alpha=0$ in Fig.~\ref{fig:friction-effect}. For oblique impacts, $\sigma_{\mathrm{peak}}$ increases gradually with $\mu$ while $\mu < \tan(|\alpha|/2)$, but once $\mu$ exceeds the trapping boundary $|\alpha| = 2\arctan\mu$, $\sigma_{\mathrm{peak}}$ rises abruptly to a high-stress plateau that then varies only weakly with further increases in $\mu$.

The key point from Fig.~\ref{fig:friction-effect} is that friction matters primarily from the standpoint of trapping the target particle and preventing ejection, but it has little role in the magnitude of $\sigma_{\mathrm{peak}}$ when a particle is trapped. In operation, the effective friction coefficient can be tuned through the surface roughness of the grinding media, coatings, or, in wet grinding, the chemistry of the slurry and its additives~\cite{iwasaki2020ball}. In these single-collision simulations, Fig.~\ref{fig:friction-effect} shows that reducing $\mu$ contracts the trapped region and can move an oblique impact from the high-stress plateau to the low-stress regime, whereas within the trapped regime $\sigma_{\mathrm{peak}}$ changes little once trapping occurs.

\subsection{Effect of Size Ratio}
\label{sec:size_ratio}

\begin{figure}[!htbp]
  \centering
  \includegraphics[width=\linewidth]{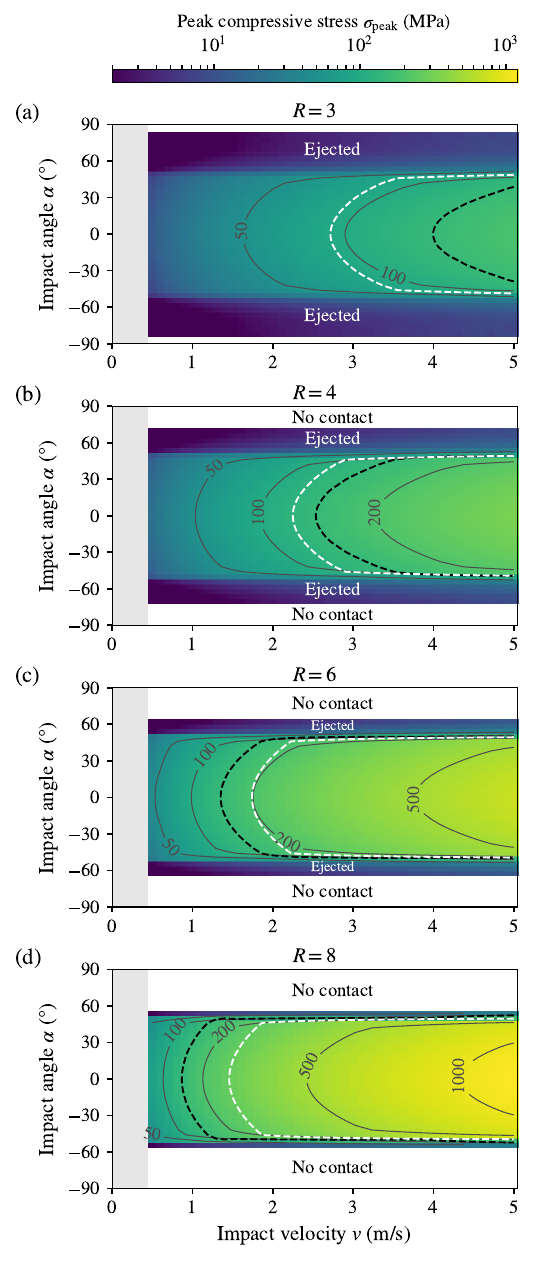}
  \caption{$v$--$\alpha$ stress maps at four values of the size ratio $R=r_g/r_t$: (a) $R=3$, (b) $R=4$, (c) $R=6$, and (d) $R=8$, all with $\gamma=0^\circ$ and $\mu=0.47$. The color, contour, and line conventions are the same as in Fig.~\ref{fig:velocity-geometry-regimes}. All four panels share a single color bar spanning $\sigma_{\mathrm{peak}}=2$--$1200$~MPa (logarithmic).}
  \label{fig:size-ratio-maps}
\end{figure}

With the context developed to this point regarding the stress in the target particle, we can examine the impact of the relative sizes of the grinding ball and target particle. Figure~\ref{fig:size-ratio-maps} shows $\sigma_{\mathrm{peak}}$ maps in the $v$--$\alpha$ plane at $\gamma=0^\circ$ and $\mu=0.47$ for four size ratios, $R=3,4,6,$ and $8$, illustrating how size ratio changes both the loading level within the trapping regime and the extent of the impact domain. As $R$ increases (smaller target particle), the peak compressive stress in the trapped portion of the parameter space rises (e.g., at $v=5$~m/s the trapped-region stress increases from about $192$~MPa at $R=3$ to $1190$~MPa at $R=8$). Consequently, smaller target particles are easier to fracture because the stress is more concentrated. At the same time, smaller particles result in the contact boundary of Equation~\eqref{eq:reach_limit} narrowing with increasing $R$ from $|\alpha|\approx 82.8^\circ$ at $R=3$ to $|\alpha|\approx 54.5^\circ$ at $R=8$, whereas the trapping boundary of Equation~\eqref{eq:trap_boundary} beyond which the target particle is ejected, $|\alpha|\approx 50.3^\circ$, is independent of $R$ at fixed $\mu$. In fact, according to Equation~\eqref{eq:trap_band}, for $R\gtrsim 9.5$ the contact boundary further narrows to lie within the trapping boundary, and the ejection region disappears.

 In terms of the general application to ball milling, finer target particles are easier to break once trapped but harder to confine, because the contact boundary of Equation~\eqref{eq:reach_limit} narrows as $R$ grows. In practice, fracture probability is maximized by keeping a balanced ball mix, with large balls for coarse feed particles and small balls for fine feed particles~\cite{katubilwa2008effect,cho2013optimum}, and by continuously removing fine particles through classification~\cite{muketekelwa2017investigating,dundar2020investigating}.

\section{Fracture Prediction}
\label{sec:fracture}

Fracture can occur only when the target particle is trapped and compressed to the point where cracks form, propagate across the particle, and split the whole target particle into a few smaller pieces, rather than just chipping the target particle near the contact points \cite{GHADIRI20005445,ghadiri2002impact,ghadiri2011aspirin,chagas2019surface}. Although the DEM simulations do not model fracture directly, we can predict fracture using two different fracture criteria.
To characterize the global failure of the target particle, the Christensen fracture criterion \cite{christensen1997yield} is applied to the stress tensor averaged over the entire particle, assuming homogeneity of the material and ignoring stress variations within the particle. Alternatively, random crack development due to the local tensile stress inside the particle under compression can be considered in terms of Weibull's weakest link theory \cite{weibull1939statistical,petrovic1987weibull} relating the fracture of the target particle to the standardized compression tests (note that Weibull's 1951 paper on statistical distributions \cite{weibull1951statistical} is sometimes incorrectly cited as the basis for Weibull's statistical theory for material strength and fracture).
The two criteria    interrogate the loaded particle at different scales, through different failure mechanisms, and with independent material inputs.

\subsection{Christensen Criterion: Failure due to Globally Averaged Stress}
\label{sec:christensen}

The Christensen failure criterion~\cite{christensen1997yield} is a relatively simple model for predicting both ductile and brittle failure in homogeneous, isotropic materials based on only two material parameters: the uniaxial tensile strength $\sigma_T$ and the compressive strength $\sigma_C$ of the material. The failure function is
\begin{equation}
  f_C = \frac{1}{k_C^2}\!\left[\frac{a_C\,k_C}{\sqrt{3}}\,\sigma_{kk} + \frac{1+a_C}{2}\,s_{ij}s_{ij}\right],
  \label{eq:fC}
\end{equation}
where $\sigma_{kk} = \sigma_{xx}+\sigma_{yy}+\sigma_{zz}$ is the trace of the stress tensor $\bar{\boldsymbol{\sigma}}$ and is related to the hydrostatic pressure, $s_{ij} = \sigma_{ij} - \tfrac{1}{3}\sigma_{kk}\delta_{ij}$ is the deviatoric stress, and $s_{ij}s_{ij} = s_{xx}^2 + s_{yy}^2 + s_{zz}^2 + 2(s_{xy}^2 + s_{xz}^2 + s_{yz}^2)$ corresponds to the square of the equivalent shear stress. Failure is predicted for $f_C \ge 1$. The two material parameters, $a_C$ and $k_C$, are determined by requiring $f_C = 1$ at the two uniaxial failure states $\sigma_{xx} = -\sigma_C$ and $\sigma_{xx} = \sigma_T$, for compression and tension respectively, with all other stress components zero, which gives:
\begin{equation}
  k_C = \frac{\sigma_C}{\sqrt{3}}, \quad\text{and}\quad
  a_C = \frac{\sigma_C}{\sigma_T} - 1.
  \label{eq:cali1}
\end{equation}

We consider brittle materials here, where $\sigma_T < \sigma_C$, which yields $a_C > 0$. For the example of limestone used here, $\sigma_C=\ensuremath{147\,\mathrm{MPa}}$ and $\sigma_T=\ensuremath{22.4\,\mathrm{MPa}}$ is the quasi-static mean tensile failure stress from three-point bend tests~\cite{grange2008dynamic}; these give $k_C \approx \ensuremath{85\,\mathrm{MPa}}$ and $a_C \approx 5.56$. $f_C$ is evaluated from the volume-averaged stress tensor of the target particle, $\bar{\boldsymbol{\sigma}}$, from LAMMPS (Equation~\eqref{eq:branch_rescale}), and therefore describes the bulk response of the particle to the imposed stress state rather than any local field within it. The failure mechanism it encodes is a global critical state: failure occurs when the particle, taken as a whole, reaches a limiting combination of hydrostatic and deviatoric stress calibrated against the uniaxial compressive and tensile strengths of the material.

The Christensen fracture boundary based on the $f_C=1$ contour line (Equation~\eqref{eq:fC}) calculated from the DEM output for $\bar{\boldsymbol{\sigma}}$ is overlaid on Figs.~\ref{fig:velocity-geometry-regimes}, \ref{fig:friction-effect}, and~\ref{fig:size-ratio-maps} (dashed black contour). As would be expected, the Christensen fracture boundary aligns with contours of $\sigma_{\mathrm{peak}}$ at low impact velocities and coincides with the trapping boundary (Equation~\eqref{eq:trap_boundary}) at higher velocities. Further note that the fracture boundary occurs at $\sigma_{\mathrm{peak}}=147$~MPa, which matches $\sigma_C=147$~MPa for the limestone material used as an example here. We explain this result in the next section.

For a head-on collision, the critical velocity based on the Christensen criterion is about $2.5$~m/s (Fig.~\ref{fig:velocity-geometry-regimes}(a,b)), and it increases as the impact and obliquity angles grow. Outside of the Christensen criterion curve, the particle is unlikely to break, while within the bounds of the curve the Christensen criterion predicts failure. In the $\alpha$--$\gamma$ plane (Fig.~\ref{fig:velocity-geometry-regimes}(c)) the non-fracture region skews toward negative $\alpha$ as $\gamma$ increases, with failure more likely at small $\alpha$ and $\gamma$, approaching a head-on collision. Since the stress decreases sharply at the trapping boundary of Equation~\eqref{eq:trap_boundary}, the Christensen criterion aligns with the trapping boundary. In Fig.~\ref{fig:friction-effect}, the Christensen failure boundary is independent of $\mu$ at $|\alpha| \approx 33^\circ$, consistent with horizontal contours of $\sigma_{\mathrm{peak}}$. For $\mu \lesssim 0.3$, where the trapping boundary $|\alpha|=2\arctan\mu$ lies within $|\alpha|\approx 33^\circ$, the Christensen failure boundary follows the trapping boundary, because the Christensen criterion is exceeded throughout the trapped region. The Christensen criterion is also shown in Fig.~\ref{fig:size-ratio-maps} for different $R$. Consistent with the higher stresses at larger $R$ and that the Christensen criterion is independent of $R$, the critical velocities predicting fracture shift progressively to lower levels for the Christensen criterion as $R$ increases.

\subsection{Christensen Criterion Based on Force}
\label{sec:christensenforce}

It is also possible to determine a critical force for failure based on the Christensen criterion without using the inferred ``per-atom'' virial tensor from LAMMPS. In other words, when the target particle is trapped in sustained double contact, the volume-averaged stress, and therefore $f_C$, can be written in closed form directly in terms of the contact forces. Under quasi-static equilibrium, this construction gives the volume-averaged compressive principal stress exactly. The approach is as follows. Since the inertia of the target particle is negligible compared with the incoming grinding ball, the target particle is in quasi-static equilibrium with $\mathbf{F}_1 +\mathbf{F}_2 \approx 0$.

Referring to Fig.~\ref{fig:trapping-condition}(b), the two forces are $\mathbf{F}_1 = F\,\hat{\mathbf{n}}_a$ and $\mathbf{F}_2 = -F\,\hat{\mathbf{n}}_a$. With the two contact points located at $\boldsymbol{\ell}_1 = -r_t\hat{\mathbf{n}}_1$ and $\boldsymbol{\ell}_2 = -r_t\hat{\mathbf{n}}_2$ from the target particle center,
the separation between the two contact points is
\begin{equation}
  \boldsymbol{\ell}_1 - \boldsymbol{\ell}_2 \;=\; -r_t\,(\hat{\mathbf{n}}_1 - \hat{\mathbf{n}}_2)  \;=\;-\,2r_t\cos\theta\;\hat{\mathbf{n}}_a,
  \label{eq:chord}
\end{equation}
where $\theta=\tfrac{1}{2}(\alpha+\gamma)$ is the half-angle between $\hat{\mathbf{n}}_1$ and $-\hat{\mathbf{n}}_2$ defined in Section~\ref{sec:trapping}.
Then Equation~\eqref{eq:avgstress} becomes
\begin{equation}
\begin{aligned}
  \bar{\boldsymbol{\sigma}}
  \;&=\; \frac{F\,\hat{\mathbf{n}}_a\otimes\boldsymbol{\ell}_1 - F\,\hat{\mathbf{n}}_a\otimes\boldsymbol{\ell}_2}{V_t}
  \;\\
  &=\; \frac{F}{V_t}\,\hat{\mathbf{n}}_a\otimes(\boldsymbol{\ell}_1-\boldsymbol{\ell}_2)
  \;=\; -\,\frac{2Fr_t\cos\theta}{V_t}\,\hat{\mathbf{n}}_a\otimes\hat{\mathbf{n}}_a .
  \end{aligned}
    \label{eq:sigma_dyad}
\end{equation}

In a frame aligned with $\hat{\mathbf{n}}_a$, the outer product $\hat{\mathbf{n}}_a\otimes\hat{\mathbf{n}}_a$ is simply $\operatorname{diag}(1,\,0,\,0)$, so
\begin{equation}
  \begin{aligned}
  \bar{\boldsymbol{\sigma}} &= \operatorname{diag}(-\sigma_0,\ 0,\ 0),\\
  \sigma_0 &= \frac{2Fr_t\cos\theta}{V_t} = \frac{3F\cos\theta}{2\pi r_t^{2}} .
  \end{aligned}
  \label{eq:uniaxial}
\end{equation}

The trapped target particle is therefore equivalently under uniaxial compression along the $n_a$ axis for every trapped geometry. In this uniaxial state, $\sigma_0$ is the magnitude of the most compressive principal stress of $\bar{\boldsymbol{\sigma}}$, so $\sigma_{\mathrm{peak}}$ (defined in Section~\ref{sec:dem}) is the maximum of $\sigma_0$ over the collision duration.

To confirm that this approach reproduces $\sigma_{\mathrm{peak}}$ from the LAMMPS virial tensor per the approach in Section~\ref{sec:dem}, the values for $\sigma_{\mathrm{peak}}$ and $\sigma_0$ are compared to the DEM cases at $v=3$~m/s, $R=4$, $\gamma=0$. Indeed, the values of $\sigma_{\mathrm{peak}}$ and $\sigma_0$ evaluated at the peak contact force agree to within 1\% for $\alpha=0$--$30^\circ$. For a head-on collision $\theta = 0$ and the uniaxial stress is a maximum; non-zero values of $\theta$ reduce the compressive stress attained at a given contact force.

Writing the normal force at either contact as $F^{\,n}=F\cos\theta$, the uniaxial compressive stress of Equation~\eqref{eq:uniaxial} becomes
\begin{equation}
  \sigma_0=\frac{3}{2\pi}\frac{F^{\,n}}{r_t^2},
  \label{eq:sigma0N}
\end{equation}
suggesting that the stress state depends primarily on the normal force and that it is the same for every trapped geometry at a given $F^{\,n}$.
For this uniaxial state, the invariants of Equation~\eqref{eq:fC} for the Christensen criterion are
\begin{equation}
  \sigma_{kk} = -\sigma_0,
  \qquad
  s_{ij}s_{ij} = \tfrac{2}{3}\sigma_0^{2},
  \label{eq:uninv}
\end{equation}
and substituting them with $k_C$ and $a_C$, as calibrated in Equation~\eqref{eq:cali1} as functions of material properties $\sigma_C$ and $\sigma_T$, gives
\begin{equation}
  f_C \;=\; 1 + \left(\frac{\sigma_0}{\sigma_C} - 1\right)\left(\frac{\sigma_0}{\sigma_T} + 1\right).
  \label{eq:fC_closed}
\end{equation}
For the trapped regime considered here, the volume-averaged stress state is uniaxial, so $f_C=1$ occurs at $\sigma_0=\sigma_C$, as noted in Section~\ref{sec:christensen}, independent of $\sigma_T$. Thus, the Christensen criterion acts here to compare the volume-averaged compressive stress to $\sigma_C$; $\sigma_T$ affects only the value of $f_C$ away from the $f_C=1$ boundary, so no fracture boundary reported here depends on $\sigma_T$. Furthermore, this volume-averaged compressive stress for the Christensen criterion does not represent the local tensile stress that splits a diametrally loaded particle, which is the failure mode against which the Weibull criterion, discussed shortly, is calibrated.

Setting $\sigma_0 = \sigma_C$ in Equation~\eqref{eq:sigma0N} yields a simple estimate of the critical normal force for failure based on the Christensen criterion.
\begin{equation}
  F_c^{\,\mathrm{C}} = \tfrac{2}{3}\pi r_t^2\sigma_C.
  \label{eq:FcC}
\end{equation}
where the subscript $c$ refers to critical and the superscript $\mathrm{C}$ refers to the Christensen criterion. $F_c^{\,\mathrm{C}}$ is called critical because it is the threshold normal force at which the Christensen function first reaches $f_C = 1$.

Thus, it is possible to determine if a collision exceeds the Christensen criterion for failure in terms of the contact forces alone without calculating the stresses in the target particle. When $F^{\,n} > F_c^{\,\mathrm{C}}$, $f_C > 1$ and the Christensen criterion is exceeded, predicting that the target particle will break. While we use the LAMMPS virial tensor approach in Section~\ref{sec:dem} to determine peak stress contours for the Christensen criterion in the stress maps in Figs.~\ref{fig:velocity-geometry-regimes}--\ref{fig:size-ratio-maps}, the approach based on the contact force is more direct and more easily implemented from the contact forces on the target particle determined from DEM simulations.

\subsection{\texorpdfstring{Weibull Criterion: Random Crack Development due to Local Stress}{Weibull Criterion: Random Crack Development due to Local Stress}}
\label{sec:weibull}

Up to this point, we have considered the overall compressive stresses in the target particle, but at a detailed level, a particle fractures under compression because of crack propagation inside it due to the compression-induced local tensile stress \cite{hiramatsu1966determination,jaeger1967failure}. Larger particles are more likely to initially contain more and larger random cracks, or flaws, and therefore, break more easily \cite{weibull1939statistical,petrovic1987weibull}.
Correspondingly, in single particle crushing tests, a characteristic compressive stress at failure $\sigma_{pc}$, also known as the crushing strength, is defined as the failure load $F_c$ normalized by the square of the particle diameter \cite{mcdowell2000application},
\begin{equation}
  \sigma_{pc} \equiv \frac{F_c}{d_p^2},
  \label{eq:sigma_pc_def}
\end{equation}
and the peak internal tensile stress driving fracture in diametral compression is proportional to $F_c/d_p^2$ through a material- and geometry-dependent numerical factor of order one \cite{hiramatsu1966determination,jaeger1967failure}. In these tests, $\sigma_{pc}$ increases as the particle diameter $d_p$ decreases \cite{mcdowell2000application,nakata2001one,qing2022effects,wang2023size}, consistent with the expected dependence on the underlying flaw population.
Weibull's weakest-link statistics \cite{weibull1939statistical,petrovic1987weibull} can be applied to explain this size dependence \cite{mcdowell2000application,nakata2001one,qing2022effects,wang2023size}, giving the scaling
\begin{equation}
  \sigma_{pc}(d_p) = \sigma_{pc,0}\left(\frac{d_p}{d_0}\right)^{-3/m},
  \label{eq:weibull_scaling}
\end{equation}
where $m$ is the Weibull modulus.
Although $\sigma_{pc,0}$ is the stress at which particles with the reference diameter $d_0$ have a $1/e$ (about $37\%$) probability of surviving the crush in the probabilistic model~\cite{mcdowell2000application}, for simplicity, we take $\sigma_{pc,0}$ as the crushing strength at which a particle fractures.
The Weibull modulus, the crushing strength, and the reference diameter are based on crushing-test experiments. For the example of a target particle made of limestone, the Weibull constants $m = 2.73$, $d_0 = \ensuremath{120\,\mathrm{mm}}$, and $\sigma_{pc,0} = \ensuremath{5.579\,\mathrm{MPa}}$ are taken from recent particle crushing tests, although these values depend on the particle size~\cite{wang2023size}. Also note that the material and testing approaches on which $\sigma_C$ and $\sigma_T$ in Table~\ref{tab:materials} are based~\cite{grange2008dynamic} are somewhat different than those for the Weibull constants~\cite{wang2023size}.

Rearranging Equation~\eqref{eq:sigma_pc_def} for the force and substituting the target particle diameter $d_p = 2r_t = 2r_g/R$ into Equation~\eqref{eq:weibull_scaling} gives the size-dependent critical force for the Weibull criterion
\begin{equation}
  F_c^{\,\mathrm{W}}(R) = \sigma_{pc}(d_p)\,d_p^2 = 4\,\sigma_{pc,0}\left(\frac{2r_g}{R d_0}\right)^{-3/m}\frac{r_g^2}{R^2}.
  \label{eq:FcW}
\end{equation}

The Weibull failure function evaluates fracture directly from the contact forces, rather than by comparing a stress tensor with material properties for the Christensen criterion.  The Weibull criterion considers breakage at the scale of the flaw population inside the particle. It is evaluated from the contact forces transmitted through the particle and is compared with the crushing force measured in single-particle diametral compression. In that configuration the particle does not fail by compression itself: the compressive load generates a transverse tensile stress along the loading axis, and it is this compression-induced tension that splits the particle apart \cite{hiramatsu1966determination,jaeger1967failure}. The strength resisting it is set by the largest flaw the particle happens to contain, so it is governed by weakest-link statistics and depends on particle size and material properties.  The material input is the characteristic crushing stress measured experimentally in single-particle crushing tests for the material of interest, limestone in this case, although the approach can be used for any material provided that crush-test results are available.

 A typical crushing test loads a particle diametrally between two platens, similar to how trapped target particles are loaded between the two grinding balls in Fig.~\ref{fig:geometric-setup} for head-on collisions. For oblique collisions, we take the compression to act along the $n_a$ axis, i.e.\ along the line connecting the two contact points.

As in Section~\ref{sec:christensenforce}, the target particle is in quasi-static equilibrium with $\mathbf{F}_1=F\,\hat{\mathbf{n}}_a$ and $\mathbf{F}_2=-F\,\hat{\mathbf{n}}_a$, where $F$ is the magnitude of the resultant contact force along the $n_a$ axis. In the DEM, the two contact-force components along the $n_a$ axis are not exactly equal because of the target particle's inertia, so we evaluate $F$ as their mean; the two agree to within $1.4\%$ at peak compression inside the trapping regime. The Weibull failure function is then defined as
\begin{equation}
f_W=\frac{F}{F_c^{\,\mathrm{W}}(R)}.
\label{eq:fW}
\end{equation}
{\emergencystretch=2em The Weibull survival probability is
$\exp[-(\sigma/\sigma_{pc})^m]$ \cite{weibull1939statistical}, which can be expressed in terms of the forces as $\exp[-(F/F_c^{\,\mathrm{W}})^m]\allowbreak=\exp(-f_W^m)$. Hence, the fracture probability is
\begin{equation}
P_f = 1 - \exp(-f_W^m).
\label{eq:Pf}
\end{equation}
Thus, $f_W=1$ corresponds to a fracture probability of $P_f = 1 - 1/e \approx 63\%$; for $m=2.73$, a fracture probability of $P_f=90\%$ corresponds to $f_W=1.36$.\par}

At this point, the Weibull fracture boundary can be determined for conditions in Section~\ref{sec:results}. Each DEM simulation yields a time series over the contact window, from which $F(t)$ is extracted and used to evaluate $f_W(t)$ via Equation~\eqref{eq:fW}.  The Weibull fracture boundary based on $f_W=1$ corresponds to $\sigma_{\mathrm{peak}}\approx 115$--$128$~MPa (128~MPa for head-on collisions with $R=4$) for a $63\%$ probability of fracture and is overlaid on Figs.~\ref{fig:velocity-geometry-regimes}, \ref{fig:friction-effect}, and~\ref{fig:size-ratio-maps}. The Weibull fracture boundary is qualitatively similar to that for the Christensen criterion, except shifted slightly to lower impact velocities in Fig.~\ref{fig:velocity-geometry-regimes}(a,b) and to larger impact angles in Fig.~\ref{fig:friction-effect}. Of course, using the $90\%$ probability instead of the $63\%$ probability would shift the Weibull criterion contour to $\sigma_{\mathrm{peak}}\approx 158$--$174$~MPa (174~MPa for head-on collisions with $R=4$). Thus, the Christensen criterion falls between the $63\%$ and $90\%$ Weibull criteria. For head-on impact the Weibull DEM critical velocity is about $2.2$~m/s, for $63\%$ fracture probability, only slightly lower than the Christensen value of about $2.5$~m/s. Like the Christensen boundary, the Weibull boundary reflects the sharp change associated with lower stresses in the ejection region.

\subsection{Comparing the Christensen and Weibull Criteria}
\label{sec:compare}

\begin{figure}[!htb]
  \centering
  \includegraphics[width=\linewidth]{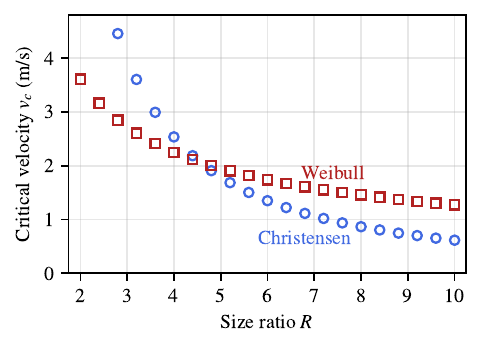}
  \caption{Head-on ($\alpha=\gamma=0^\circ$) critical velocity $v_c$ versus size ratio $R=r_g/r_t$ for $R=2$--10 at $\mu=0.47$, with $r_g=25$~mm fixed and $r_t$ varied; open blue circles, Christensen criterion ($f_C=1$), and open red squares, Weibull criterion ($f_W=1$). Conditions under which no fracture occurs for $v<5$~m/s are not shown.}
  \label{fig:vc-R}
\end{figure}

Differences between the Christensen and Weibull criteria become evident in Fig.~\ref{fig:size-ratio-maps} with changing $R$ in that the ordering of the two criteria reverses between $R=4$ and $R=6$: the Christensen boundary corresponds to higher velocity than the Weibull boundary at $R=3$ and $4$, but corresponds to lower velocity for $R=6$ and $8$. To more clearly show the dependence of the two criteria on $R$, Fig.~\ref{fig:vc-R} compares the head-on critical velocities measured from DEM simulations as $R$ is varied from 2 to 10 for head-on collisions ($\alpha=\gamma=0^\circ$) at $\mu=0.47$. Impact velocities from 0.5 to 5~m/s (in steps of 0.1~m/s) are simulated, and the critical velocity is taken as the velocity at which the corresponding failure function first reaches 1, interpolated linearly between the simulated velocities. The resulting critical velocity curves cross at $R\approx 4.6$, consistent with the reversal between the Christensen and Weibull boundaries between $R=4$ and $R=6$ in Fig.~\ref{fig:size-ratio-maps}. This crossover can also be found analytically using Equations~\eqref{eq:FcC} and~\eqref{eq:FcW}, noting that it occurs at $F_c^{\,\mathrm{C}} = F_c^{\,\mathrm{W}}$. The Weibull critical velocity decreases more gently with $R$ because the size effect of Equation~\eqref{eq:weibull_scaling} makes smaller particles statistically stronger, partially offsetting the increase in stresses for smaller target particles.

It is evident that Equation~\eqref{eq:sigma0N} for the Christensen criterion, $\sigma_0=(3/2\pi) F^n/ r_t^2$, and Equation~\eqref{eq:sigma_pc_def} for the Weibull criterion, $\sigma_{pc} = (1/4)F_c/r_t^2$, are similar in form. However, they differ in both their prefactors and the forces that are considered. The Christensen criterion is based on the normal force at the contacts, while the statistical Weibull criterion is based on common crushing-test expressions for a characteristic failure stress~\cite{mcdowell2000application,wang2023size}. An alternative approach to the estimate for the peak local tensile stress on the loading axis is expressed as $(\kappa/\pi) F/ r_t^2$ with $0.7\leq\kappa\leq 1.4$ ($0.22\leq\kappa/\pi\leq0.45)$~\cite{hiramatsu1966determination,jaeger1967failure}. The coefficient $3/2\pi~(= 0.48)$ in Equation~\eqref{eq:sigma0N} arises because $\sigma_0$ is the volume-averaged axial compressive stress over the entire target particle for two opposite contact forces (with zero volume-averaged transverse components), whereas crushing-test formulas are based on stress normalizations or peak tensile stresses that drive target particle breakage as treated through the Weibull criterion using $\sigma_{pc}$. Thus, the Christensen and Weibull criteria impose thresholds on different components of the same contact force: the Christensen threshold $F_c^{\,\mathrm{C}}$ (Equation~\eqref{eq:FcC}) applies to the normal component at either contact, $F^{\,n}=F\cos\theta$, whereas the Weibull threshold $F_c^{\,\mathrm{W}}(R)$ (Equation~\eqref{eq:FcW}) applies to the resultant $F$ along the $n_a$ axis, as in a diametral crushing test.

\section{Conclusions}\label{sec:conclusions}

This study combines DEM simulations, two independent fracture criteria, and analytic contact geometry to connect three-body collision mechanics to particle fracture in ball milling. A target particle initially placed between a fixed and an incoming grinding ball is simulated over wide ranges of impact velocity, impact angle, obliquity angle, size ratio, and friction coefficient, and each collision is evaluated with a stress-based Christensen criterion and a force-based Weibull criterion. Across this parameter space, two governing mechanisms emerge: a geometry-based regime, bounded by the closed-form trapping condition, in which friction and collision orientation determine whether the target particle is trapped at all, and a velocity-based regime, in which the impact velocity determines whether a trapped particle reaches the fracture threshold of either criterion.

For fixed sizes and friction, the most effective fracture occurs in head-on collisions at sufficiently high velocity. Within the trapping regime the critical velocity is lowest for head-on impacts and rises as the collision angles grow; outside the trapping regime, the collision is glancing, the particle is ejected, and the outcome becomes independent of velocity. Finer particles fracture more easily once trapped but are less likely to remain confined, and friction mainly determines the width of the trapping regime while having a much weaker effect on the load inside it. The Christensen and Weibull criteria for fracture are qualitatively similar.

In practice, fracture probability is raised by steering collision kinematics toward near-head-on, low-offset impacts and by managing the size distribution of the charge. The former can be achieved by operating at rotation speeds that promote cataracting and by lifter and liner designs that favor direct impacts; the latter by keeping a balanced ball mix and continuously removing fine particles through classification. The friction that holds the trapping regime open can be tuned through the surface condition of the grinding media or, in wet grinding, the slurry and its additives. Since the trapping regime gains little beyond moderate friction, over-lubrication that shrinks the trapping regime is the main condition to avoid.

The three-sphere model for ball milling presented here is clearly a minimal model. Although we have used steel grinding balls and limestone target particles as an example here, the approach can be applied for any materials of interest. In future work, it would be reasonable to follow Professor Mojtaba Ghadiri’s approach of uncovering fundamental links between particle-scale interactions and bulk particle behavior in real systems. For instance, one could simulate a computationally feasible slice of a more realistic ball mill including lifters and representative fill, and extract the joint distribution of impact conditions experienced by the target particles. Convolving that distribution with fracture maps like those developed here would yield mill-scale predictions of how often particles are trapped and squeezed, and of the aggregate breakage probability at a given operating point. A particle replacement model \cite{chagas2019development,chagas2021stochastic} can then be embedded, in which a particle satisfying the fracture criterion is replaced by daughter fragments sampled from a measured breakage function \cite{BOTTCHER2021554}, connecting the single-collision physics established here to measured size-reduction trajectories. While grinding operations consume a large share of industrial energy with only a very small fraction of that energy resulting in comminution, identifying the collisions that actually break particles is the first step toward spending that energy more effectively.

\section*{Declaration of\\Competing Interest}
The authors declare that they have no known competing financial interests or personal relationships that could have appeared to influence the work reported in this paper.

\section*{Acknowledgments}
The authors thank Paul B. Umbanhowar of Northwestern University for helpful feedback, and Anderson S. das Chagas of NETZSCH-Feinmahltechnik GmbH for suggesting this line of research.

\section*{Data Availability}
The simulation data (sweep\_results.csv), LAMMPS input templates, and analysis and plotting scripts are openly available on Zenodo~\cite{cui2026zenodo} at \href{https://doi.org/10.5281/zenodo.23633121}{\nolinkurl{https://doi.org/10.5281/zenodo.23633121}}. The deposit also includes raw LAMMPS output for the illustrative runs used in Fig.~\ref{fig:outcomes} and in Fig.~\ref{fig:si_dump} of the Supplementary Information.

\section*{Author Contributions}
\textbf{Ziyi Cui:} Conceptualization, Methodology, Software, Formal analysis, Visualization, Writing -- original draft.
\textbf{Qiong Zhang:} Conceptualization, Supervision, Methodology, Writing -- review and editing.
\textbf{Suwei Liu:} Supervision, Methodology, Writing -- review and editing.
\textbf{Richard M. Lueptow:} Conceptualization, Supervision, Methodology, Writing -- review and editing.

\section*{Declaration of generative AI and AI-assisted technologies in the manuscript preparation process}
During the preparation of this work, the authors used Claude (Anthropic), accessed through Claude Code, in order to debug the simulation-setup and post-processing scripts and figures and to check derivations and numerical values, and GPT-5.2 (OpenAI) in order to edit the manuscript text. After using these tools/services, the authors reviewed and edited the content as needed and take full responsibility for the content of the published article.

\FloatBarrier
\bibliographystyle{elsarticle-num1}
\bibliography{references}
\clearpage
\onecolumn
\small
\begin{center}
  {\Large\bfseries Supplementary Information}\\[6pt]
  {\large Trapping and Fracture of Particles in Three-Sphere Ball-Milling Collisions}\\[6pt]
  Ziyi Cui, Qiong Zhang, Suwei Liu, and Richard M.\ Lueptow
\end{center}
\vspace{8pt}

\noindent
Equation, figure, and section numbers prefixed with S refer to this document;
unprefixed numbers refer to the main text.

\vspace{6pt}

\renewcommand{\thesection}{S\arabic{section}}
\setcounter{section}{0}
\renewcommand{\theequation}{S\arabic{equation}}
\setcounter{equation}{0}
\renewcommand{\thefigure}{S\arabic{figure}}
\setcounter{figure}{0}
\renewcommand{\thetable}{S\arabic{table}}
\setcounter{table}{0}
\renewcommand{\theHequation}{S\arabic{equation}}
\renewcommand{\theHsection}{S\arabic{section}}
\renewcommand{\theHfigure}{S\arabic{figure}}
\renewcommand{\theHtable}{S\arabic{table}}

\section{Nomenclature}\label{sec:si_nomenclature}
Symbols used in the main text and the Supplementary Information.

\begin{longtable}{@{}L{2.0cm} L{\dimexpr\linewidth-2.0cm-2\tabcolsep\relax}@{}}
\toprule
Symbol & Meaning \\
\midrule
\endfirsthead
\toprule
Symbol & Meaning \\
\midrule
\endhead
\bottomrule
\endfoot

$\gamma$
& Obliquity angle, between the incoming grinding ball's velocity and the $n_2$ axis.\\

$\alpha$
& Impact angle, between $-\mathbf{v}$ and the $n_1$ axis, as in the main text.\\

$\theta$
& Contact half-angle, $\theta = \tfrac12(\alpha+\gamma)$. The contact and trapping conditions depend on the two collision angles
  only through $\theta$. \\

$R$
& Size ratio, $R = r_g/r_t$, of grinding-ball to target-particle radius. \\

$r_g$, $r_t$
& Radii of the grinding balls and the target particle; subscripts $g$ and $t$ denote the grinding balls and the target particle throughout. \\

$\mathbf{C}_f$, $\mathbf{C}_i$, $\mathbf{C}_t$
& Centers of the fixed grinding ball, the incoming grinding ball, and the target particle. \\

$n_1$, $n_2$, $n_a$
& Axes: $n_1$ passes through $\mathbf{C}_f$ and $\mathbf{C}_t$, $n_2$ through $\mathbf{C}_i$ and $\mathbf{C}_t$, and $n_a$ is the common line of action of the two resultant contact forces of a trapped target particle. \\

$\mathbf{F}_i$, $F_i^{\,n}$, $F_i^{\,t}$
& Contact force at contact $i$ ($i=1$ fixed grinding ball, $i=2$ incoming grinding ball) and its normal and tangential components, along the unit vectors $\hat{\mathbf{n}}_i$ and $\hat{\mathbf{t}}_i$. \\

$\mu$
& Sliding friction coefficient; $\mu_s$ denotes the same-material values in Table~\ref{tab:materials}. \\

$\phi$
& Half-angle of the Coulomb friction cone, $\phi = \arctan\mu$. The trapping
  condition is $|\theta| \le \phi$. \\

$\rho$
& Mass density. \\

$\nu$
& Poisson's ratio. \\

$E$
& Young's modulus. \\

$e$
& Normal coefficient of restitution of the limestone--steel contact. \\

$\bar{\boldsymbol{\sigma}}$, $\sigma_{ij}$
& Average (virial) stress tensor in the target particle, with trace
  $\sigma_{kk}$ and deviator $s_{ij}$. \\

$\sigma_C$, $\sigma_T$
& Uniaxial compressive and tensile strengths of the target material, which
  calibrate the Christensen constants $k_C$ and $a_C$. \\

$\sigma_0$
& Magnitude of the uniaxial volume-averaged compressive stress of a trapped target particle along the $n_a$ axis. \\

$V_t$
& Target particle volume, $V_t = \tfrac{4}{3}\pi r_t^3$. \\

$\mathbf{r}_i$
& Center-to-center vector from the target particle to the grinding ball at contact $i$, used in the LAMMPS per-atom virial stress $\bar{\boldsymbol{\sigma}}_{\mathrm{LAMMPS}}$. \\

$\kappa$
& Numerical factor in the estimate $(\kappa/\pi)F/r_t^2$ of the peak local tensile stress on the loading axis. \\

$\sigma_{pc}$, $\sigma_{pc,0}$
& Characteristic particle crushing stress and its value at the reference diameter
  $d_0$, from single-particle crushing tests (Weibull criterion). \\

$d_p$, $d_0$
& Particle diameter in the Weibull size scaling ($d_p = 2r_t$ for the target particle) and the reference diameter. \\

$m$
& Weibull modulus. \\

$P_f$
& Fracture probability, $P_f = 1-\exp(-f_W^m)$. \\

$\delta$
& Normal contact overlap: $\delta_1$, $\delta_2$ at contacts~1 and~2. \\

$\delta_{ij}$
& Kronecker delta, in the deviatoric-stress definition of the Christensen
  criterion. \\

$\Delta t$
& Integration timestep of the velocity-Verlet scheme,
  $1.0\times10^{-7}\,$s. \\

$v$ & Impact velocity of the incoming grinding ball. \\

$v_c$ & Critical impact velocity for fracture. \\

$\mathbf{v}$ & Initial in-plane velocity of the incoming grinding ball, of magnitude $v$. \\

$t$, $i$ & Time; contact index ($i=1,2$). \\

$g$, $D$ & Gravitational acceleration and mill inside diameter (Section~\ref{sec:si_ranges}). \\

$\sigma_{\mathrm{peak}}$ & Peak compressive principal stress of the target particle over the contact window (Section 3 of the main text). \\

$f_C$, $f_W$ & Christensen and Weibull failure functions; fracture is predicted when either reaches one. \\

$F_c^{\,\mathrm{C}}$, $F_c^{\,\mathrm{W}}$ & Critical contact forces of the Christensen and Weibull criteria. \\

$F$, $F^{\,n}$ & Magnitude of the resultant contact force on the target particle along the $n_a$ axis, which enters the Weibull failure function, and its normal component at either contact, $F^{\,n}=F\cos\theta$. \\

$\boldsymbol{\ell}_i$
& Branch vector from the target-particle center to contact $i$, in the
  average-stress formula. \\

$\pi$
& The circle constant. \\

\end{longtable}

\clearpage
\section{Velocity and size ranges of the simulations}\label{sec:si_ranges}
The velocity range of the simulations, 0.5--5~m/s, spans the shell speeds of typical ball mills, which set the scale of the grinding-media velocities: the 25.4~cm laboratory batch mill of Datta and Rajamani~\cite{datta2002direct} turns at 55~rpm, a shell speed of 0.73~m/s, while the 5~m industrial mill of Cleary~\cite{cleary2001charge}, 4.8~m inside the liners, has a critical speed of 19.5~rpm, giving a shell speed of 4.9~m/s at the critical speed and 3.7~m/s at the 75\% of critical, typical of operation. A grinding ball falling freely across the 4.8~m inside diameter ($D$) of that mill would reach $\sqrt{2gD}\approx 9.7$~m/s; this is an upper bound on the impact velocity of a cataracting ball, and the simulated range lies below it. The grinding ball and target particle sizes are drawn from the same setting: the charge of that mill consists of 50--100~mm steel balls and 5--50~mm rock, against 50~mm balls and 5--25~mm target particles here.

\section{Numerical convergence of the recording interval}
\label{sec:si_convergence}

\begin{figure}[!htb]
  \centering
  \includegraphics[width=0.96\linewidth]{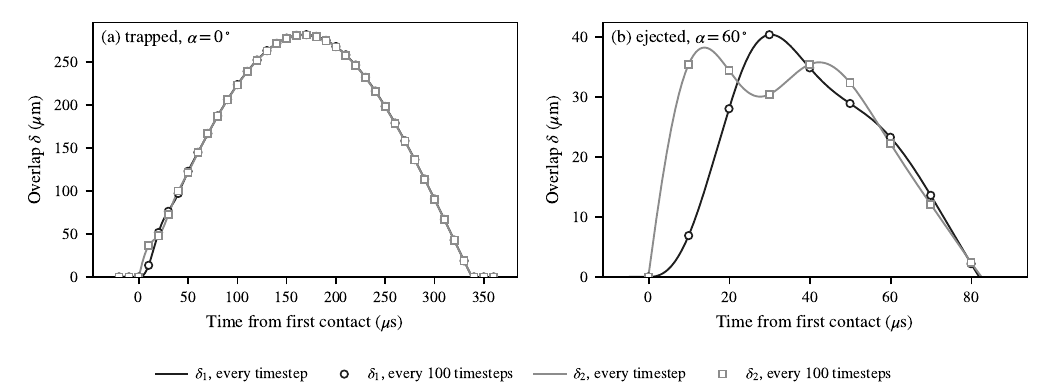}
  \caption{Deformation histories of the target particle expressed as the normal overlaps $\delta_1$ and $\delta_2$ at contacts 1 and 2, recorded at every timestep and at the production interval of every $100$ timesteps. In the simulation, a steel grinding sphere of radius 25~mm collides with a limestone target particle of radius 6.25~mm (size ratio $R = 4$) at $v = 5$~m/s with friction coefficient $\mu = 0.47$ and obliquity angle $\gamma = 0^\circ$. Panel (a) shows the trapped head-on case ($\alpha = 0^\circ$) and panel (b) shows the ejected case ($\alpha = 60^\circ$). The black solid line denotes $\delta_1$ recorded at every timestep and the black open circles denote $\delta_1$ subsampled every $100$ timesteps, while the gray solid line denotes $\delta_2$ recorded at every timestep and the gray open squares denote $\delta_2$ subsampled every $100$ timesteps. The horizontal axis is time from first contact in $\mu$s and the vertical axis is overlap in $\mu$m.}
  \label{fig:si_dump}
\end{figure}

The failure functions are evaluated from simulation output recorded every $100$ timesteps, and the recorded extrema can in principle underestimate the true force and stress peaks if the maximum occurs between two stored outputs; for three head-on collisions at $v = 0.95\,v_c$, $v_c$, and $1.05\,v_c$ ($R = 4$, $\mu = 0.47$), $\sigma_{\mathrm{peak}}$ and the peak $f_W$ re-evaluated at recording intervals $N = 1$ to $400$ were converged to within $0.13\%$ at the production interval $N = 100$, and the head-on critical velocity changed by only $0.09\%$. To provide a direct, time-resolved demonstration at the maximum impact speed, Figure~\ref{fig:si_dump} reports the deformation histories for two additional simulations at $v = 5$~m/s, $R = 4$, $\mu = 0.47$, and $\gamma = 0^\circ$, in which the normal overlaps $\delta_1$ and $\delta_2$ at contacts 1 and 2 are recorded at every timestep ($10^{-7}$~s) and, using the same trajectories, at the production interval of every $100$ timesteps. In the trapped case at $\alpha = 0^\circ$, the target particle remains pinned between the balls for a $338$~$\mu$s contact ($34$ recorded points at the production interval), and both overlaps rise and fall together along a single compression--rebound arc such that $\delta_1$ and $\delta_2$ overlay almost exactly, consistent with both contacts carrying essentially the same load. The peak overlap in this trapped collision is $281.7$~$\mu$m in the every-timestep record and $281.6$~$\mu$m in the every-$100$-timestep record, showing negligible sensitivity to the recording interval under sustained, symmetric loading. In the ejected case at $\alpha = 60^\circ$, which lies beyond the trapping limit $2\arctan\mu = 50.3^\circ$ while remaining within the contact limit of $73.7^\circ$, the collision persists for $82.5$~$\mu$s (eight recorded points at the production interval) and exhibits an initial strike by the incoming ball for which $\delta_2$ rises to $38.2$~$\mu$m within $14$~$\mu$s before the target particle is driven into the fixed ball and $\delta_1$ reaches $40.4$~$\mu$m at $30$~$\mu$s while sliding laterally. During this ejected collision, $\delta_2$ dips and recovers once while the incoming ball stays in contact with the target particle, after which the target particle exits the gap at $7.2$~m/s and both overlaps return to zero at $82.5$~$\mu$s; the every-$100$-timestep record reproduces the overall trajectory and duration while underestimating the first, brief peak of $\delta_2$ by $7\%$ (recovered peaks $40.4$~$\mu$m for $\delta_1$ and $35.5$~$\mu$m for $\delta_2$). In this ejected collision, the peak compressive stress remains approximately $10$ times below the compressive strength ($\sigma_{\mathrm{peak}} = 15.2$~MPa vs.\ $\sigma_C = 147$~MPa), so the modest undersampling of the earliest $\delta_2$ peak does not affect any fracture evaluation (Christensen failure function $f_C = 0.026$).

\FloatBarrier
\end{document}